\documentclass[aps,pra,showpacs,groupedaddress,superscriptaddress,onecolumn]{revtex4}

\usepackage{graphicx}
\usepackage{amsmath}

\input{tcilatex}

\begin{document}

\title{Continuous photodetection model: quantum jump engineering and hints
for experimental verification}
\author{A. V. Dodonov}
\email{adodonov@df.ufscar.br}
\author{S. S. Mizrahi}
\email{salomon@df.ufscar.br} \affiliation{Departamento de
F\'{\i}sica, CCET, Universidade Federal de S\~{a}o Carlos, Via
Washington Luiz km 235, 13565-905, S\~ao Carlos, SP, Brazil}
\author{V. V. Dodonov}
\email{vdodonov@fis.unb.br} \affiliation{Instituto de F\'{\i}sica,
Universidade de Bras\'{\i}lia, PO Box 04455, 70910-900,
Bras\'{\i}lia, DF, Brazil}
\date{\today }

\begin{abstract}
We examine some aspects of the continuous photodetection model for
photocounting processes in cavities. First, we work out a
microscopic model that describes the field-detector interaction and
deduce a general expression for the Quantum Jump Superoperator
(QJS), that shapes the detector's post-action on the field upon a
detection. We show that in particular cases our model recovers the
QJSs previously proposed ad hoc in the literature and point out that
by adjusting the detector parameters one can engineer QJSs. Then we
set up schemes for experimental verification of the model. By taking
into account the ubiquitous non-idealities, we show that by
measuring the lower photocounts moments and the mean waiting time
one can check which QJS better describes the photocounting
phenomenon.
\end{abstract}

 \maketitle

\maketitle

\section{Introduction}

The subject of quantum measurements is as old as the very foundation of
quantum mechanics. For a long time the scheme of sudden state reduction,
proposed by von Neumann, has been prevalent. He conjectured that the
measurement of an observable on a system entails its state reduction to one
of its eigenstates, or shortly, a sudden change of the system state by
projection. However, on probing an electromagnetic (EM) field state through
a photocount process, the photons are detected and counted one by one, a
photon entering a photomultiplier tube provokes a burst of electrons (a
photocurrent) which is viewed as originating from that single photon. It is
then registered and counted. A sequence of bursts in a given time interval
is associated to the photocount process. So, the determination of the field
state is not achieved by an instantaneous projective measurement, but it
takes some time $t$ to count a sequence of photons, whose statistics gives
information about the field state. A classical theory describing this
process was proposed by Mandel~\cite{mandel}. Quantum photocount theories
were developed by several authors \cite{glauber,mandel2,kk,mollow,scully}
(see the review~\cite{PL-rev} for more references). These theories rely on
the assumption of an instantaneous measurement of $m$ photons, independently
of the duration of the detection. However, actually, photons are counted
sequentially, one by one, and the time intervals between counts is irregular
and uncontrollable.

For describing more realistically a sequential photocount events in
an ideal closed cavity, Srinivas and Davies (SD)~\cite{SD} proposed
an approach based on the concept of continuous measurement. Their
scheme allows calculating various statistical functions, that can be
compared with experimental outcomes, such as the probability for
counting any number $k$ of photons in a time interval $t$ and
different coincidence probability densities. The SD model takes into
account a \textquotedblleft back action\textquotedblright\ of the
photodetector on the state of field and gives the conditioned field
state, i.e., the field state just after a given sequence of
photocounting events. A progress in understanding the physical
meaning of the axiomatic SD model was achieved due to studies
\cite{milburn1,ueda1,ueda2,imoto,ueda4} (for other references see
\cite{PL-rev,carmichael}).

Continuous photodetection model (CPM) is extensively discussed in
the literature \cite{PL-rev,milburn1,ueda2,SD1,OMD-JOB}, so we shall
mention only its main properties. The model, also referred as a
theory, describes the field state evolution during the
photodetection process in a closed cavity and is formulated in terms
of two fundamental \textit{operations}, assumed to represent the
\textit{only} events occurring at each infinitesimal time interval.
(1) The one-count operation, represented by the \textit{Quantum Jump
Superoperator} (QJS), describes the detector's back-action on the
field upon a single count, and the trace calculation over the QJS
gives the probability per unit time for occurrence of a detection.
(2) The \textit{no-count} operation describes the field non-unitary
evolution in absence of counts.

If one sets the formal expressions for these operations, all possible
outcomes of a photocounting experiment can be predicted. For instance, the
photocounts \cite{SD,ueda1,ueda2} and the waiting time (WT) \cite%
{waiting1,waiting2,waiting3,waiting4} statistics are among the most
common quantities studied both theoretically and experimentally.
Moreover, CPM conferred a new step in photodetection theories by
allowing one to determine the field state after an arbitrary
sequence of measurements, thus creating the possibility of
controlling the field properties in real time experiments
\cite{a7,a10,a17}.

Actually, the QJS is the main formal ingredient within the theory,
since it also dictates the form of the no-count superoperator
\cite{SD}. Two different models for the QJS were proposed \emph{ad
hoc}. The first was proposed by Srinivas and Davies (SD) in the
original paper \cite{SD} (we call it \textit{SD-model}) as
\begin{equation}
\hat{J}_{SD}\rho =\mathcal{R}\hat{a}\rho \hat{a}^{\dagger },  \label{01}
\end{equation}%
where $\rho $ is the field statistical operator, $\hat{a}$ and
$\hat{a}^{\dagger }$ are the usual bosonic annihilation/creation
operators and $\mathcal{R}$ is roughly the detector's ideal counting
rate \cite{SD,DMD-JOB05}. From the very beginning
the authors \cite{SD} denounced the presence of some inconsistences when $%
\hat{J}_{SD}$ is employed for describing a real photodetection process, this
point was also appointed in \cite{DMD-JOB05}. Nevertheless, this QJS is
widely used in the literature \cite%
{ueda2,ueda4,ueda6,a2,a4,a6,a7,a9,a10,a12,a13,a16,b222}.

The other proposal \cite{OMD-JOB,benaryeh} assumes for the QJS an expression
written in terms of the ladder operators $\hat{E}_{-}=\left( \hat{a}%
^{\dagger }\hat{a}+1\right) ^{-1/2}\hat{a}$ and $\hat{E}_{+}=\hat{E}%
_{-}^{\dagger }$ (also known as \textit{exponential phase operators} \cite%
{p1,p2,Vourd92,p3,p4})
\begin{equation}
\hat{J}_{E}\rho =\mathcal{R}\hat{E}_{-}\rho \hat{E}_{+}.  \label{02}
\end{equation}%
In \cite{DMD-JOB05}
we called \textit{E-model} such a choice, to distinguish it from the SD QJS (%
\ref{01}). We note that $\mathcal{R}$ may be different for SD- and
E- models, but the above notation will not cause confusion in this
paper. Besides eliminating the inconsistencies within the SD-model,
the use of\ the E-model leads to different qualitative and
quantitative predictions for several observable quantities.

In section \ref{sub1} we present a microscopic model for the
detector assumed to be composed of a sensor (2-level quantum object)
and an
amplification mechanism (macroscopic thermal reservoir). In section \ref%
{sub2} we compare our model's predictions concerning photodetector
properties with experimental data and show that the QJSs (\ref{01}) and (%
\ref{02}) are particular cases of a general time-dependent \textit{%
transition superoperator}, each one occurring in a particular regime
of the detector experimental parameters \cite{QJS,EQJS}. Moreover,
we point out that by manipulating detector's parameters one could
engineer the form of the QJS, thus changing the dynamics of the
photodetection, as well as the field state after a sequence of
measurements.

A way to check the validity of CPM and to decide which QJS better
describes the phenomenon in practice can be accomplished through
photodetection experiments in a high finesse cavity by comparing
experimental outcomes with theoretical predictions. However, real
detectors and cavities are far from ideal. So in section \ref{sub}
we include the main non-idealities [quantum efficiency (QE) and dark
counts] into the CPM and deduce general expressions for the
photocounts and the WT distributions. As a practical application, in
section \ref{sub3} we give some experimental hints to decide
which QJS model actually prevails in a photodetection experiment. Section %
\ref{secc} contains a summary and the conclusions.

\section{Microscopic model of 2-level photodetector}

\label{sub1}

We model the photodetector as constituted of two parts: the sensor
and the amplification mechanism (AM). The sensor is a two-level
quantum object (atom-like) with resonant frequency $\omega _{0}$,
interacting with the monomodal EM field of frequency $\omega $. It
has the ground $|g\rangle $ and the excited $|e\rangle $ states, so
we describe it by the usual Jaynes--Cummings Hamiltonian \cite{JCM}
\begin{equation}
H_{0}=\frac{1}{2}\omega _{0}\sigma _{0}+\omega {\hat{n}}+g\hat{a}\sigma
_{+}+g^{\ast }\hat{a}^{\dagger }\sigma _{-},  \label{JCH}
\end{equation}%
where $g$ (assumed to be real, since only its absolute value enters the
final expressions) is the sensor-field coupling constant, {%
and }the sensor operators are $\sigma _{0}=|e\rangle \langle
e|-|g\rangle \langle g|,$ $\sigma _{+}=|e\rangle \langle g|$ and
$\sigma _{-}=|g\rangle \langle e|$. The interaction described by
Hamiltonian (\ref{JCH}) allows a coherent exchange of excitations
between the sensor and the field -- the Rabi oscillations.

Upon absorbing a photon the sensor initially in $|g\rangle $ makes a
transition to $|e\rangle $, and after some time it decays back,
emitting a \textit{photoelectron} into the AM; after that, the
detector is ready for the next photodetection. The AM is a complex
macroscopic structure (e.g. photomultiplier tube) that somehow
amplifies the photoelectron and originates some observable
macroscopic effect, giving rise to the click of the detector. In
order to describe general features of the AM independent of the type
of the single photon detector (SPD), we model it as a macroscopic
thermal reservoir with a mean number of intrinsic excitations
$\bar{n}$ (due to the effects of temperature and internal defects).
Thus, the whole system field--SPD unconditioned time evolution (when
the detector is not monitored \cite{carmichael}) is described by the
master equation \cite{EQJS}
\begin{eqnarray}
\dot{\rho}_{T} &=&\frac{1}{i}\left[ H_{0},\rho _{T}\right] -\gamma \overline{%
n}\left( \sigma _{-}\sigma _{+}\rho _{T}-2\sigma _{+}\rho _{T}\sigma
_{-}+\rho _{T}\sigma _{-}\sigma _{+}\right)   \notag \\
&&-\gamma \left( \overline{n}+1\right) \left( \sigma _{+}\sigma _{-}\rho
_{T}-2\sigma _{-}\rho _{T}\sigma _{+}+\rho _{T}\sigma _{+}\sigma _{-}\right)
,  \label{eqmestra}
\end{eqnarray}%
where $\gamma $ is the sensor--AM coupling constant.

According to CPM, the trace of QJS applied on the field density operator
gives the probability density $\mathrm{{p}(t)}$ for the photodetection, i.e.
emission of a photoelectron at time $t$, given that at time $t=0$ the
detector-field system was in the state
\begin{equation}
\rho _{0}=|g\rangle \langle g|\otimes \rho ,  \label{rhho0}
\end{equation}%
where $\rho $ is the field statistical operator. Microscopically
this means that initially the detector is in the ground state; then,
during the time interval $(0,t)$ the sensor interacted with the
field and it could have absorbed a photon, doing a transition
$|g\rangle \rightarrow |e\rangle $. So $\mathrm{{p}(t)}\Delta t$ is
the probability of the sensor decaying back to $|g\rangle $ during
the time interval $(t,t+\Delta t)$ and simultaneously emitting a
photoelectron that will lately originate one click. Here, the
emission of the photoelectron is our interpretation of how the
detector operates, and this phenomenon does not appear explicitly in
the formalism.

Following the quantum trajectories approach \cite{carmichael}, $\mathrm{p}(t)
$ is calculated as%
\begin{equation}
\mathrm{p}(t)=\mathrm{Tr}_{F-D}\left[ \hat{R}\hat{U}_{t}\rho
_{0}\right] ,\label{ddd}
\end{equation}%
where $\hat{U}_{t}\rho _{0}$ represents the evolution of the
field-SPD system from initial state $\rho _{0}$ at time $t=0$ to the
time $t$ without detections, and $\hat{R}\hat{U}_{t}\rho _{0}$ stays
for a click (instantaneous decay of the sensor) at the time $t$ (in
the trace, F stands for the field and D -- for the detector). The
sensor instantaneous decay $|e\rangle \rightarrow |g\rangle $ is
represented by the superoperator
\begin{equation}
\hat{R}\rho _{0}=2\gamma \left( \overline{n}+1\right) \sigma _{-}\rho
_{0}\sigma _{+},  \label{decay}
\end{equation}%
whose trace gives the probability density of such an event. In Eq. (\ref%
{decay}) $\gamma $ stands for the sensor zero-temperature decay rate, and we
included the term\ $\overline{n}+1$ because it is natural to assume that the
rate of decays is proportional to the effective temperature of the detector (proportional to $\bar{n}$).
The complementary no-decay superoperator $%
\hat{U}_{t}\rho _{0}=\rho _{U}(t)$ describes the non-unitary
evolution of the field--SPD system during time interval $(0,t)$
without clicks; it is the
solution to the master equation (\ref{eqmestra}) without the decay term (\ref%
{decay}):
\begin{equation}
\frac{d\rho _{U}}{dt}=-i(H_{e}\rho _{U}-\rho _{U}H_{e}^{\dagger })+2\gamma
\overline{n}\sigma _{+}\rho _{U}\sigma _{-},  \label{rhoU11}
\end{equation}%
where
\begin{equation}
H_{e}=\frac{\left( \omega _{0}-i\gamma \right) }{2}\sigma _{0}+\omega {\hat{n%
}}+g\hat{a}\sigma _{+}+g\hat{a}^{\dagger }\sigma _{-}-i\gamma (\overline{n}+%
\frac{1}{2}).  \label{rhoU12}
\end{equation}%
Taking the partial trace over the detector variables in (\ref{ddd})
one obtains the superoperator
\begin{equation}
\hat{\Xi}(t)\rho =\mathrm{Tr}_{D}\left[ \hat{R}\hat{U}_{t}\rho _{0}\right] ,
\label{transition}
\end{equation}%
which describes the back-action of the detector on the field upon
one click -- it is the \emph{transition superoperator} \cite{QJS}
and, as will be seen below, its time average defines the QJS.
Moreover, the probability density for a count is simply
$\mathrm{p}(t)=\mathrm{Tr}\left[ \Xi (t)\rho \right] $.

In order to solve Eq. (\ref{rhoU11}) we first do the transformation%
\begin{equation}
\rho _{U}=X_{t}\tilde{\rho}_{U}X_{t}^{\dagger },\quad X_{t}\equiv \exp
(-iH_{e}t)
\end{equation}%
to obtain a simple equation for $\tilde{\rho}_{U}$
\begin{equation}
\frac{d\tilde{\rho}_{U}}{dt}=2\gamma \overline{n}\widetilde{\sigma }_{+}%
\tilde{\rho}_{U}\widetilde{\sigma }_{-},\quad \widetilde{\sigma }%
_{+}=X_{-t}\sigma _{+}X_{t},\qquad \widetilde{\sigma }_{-}=\widetilde{\sigma
}_{+}^{\dagger },
\end{equation}%
whose formal solution is
\begin{equation}
\tilde{\rho}_{U}(t)=\rho _{0}+2\gamma \overline{n}\int_{0}^{t}dt^{\prime }%
\widetilde{\sigma }_{+}(t^{\prime })\tilde{\rho}_{U}(t^{\prime })\widetilde{%
\sigma }_{-}(t^{\prime }).  \label{sasa}
\end{equation}

If one iterates Eq. (\ref{sasa}) and substitutes the result into Eq. (\ref%
{transition}) one gets the following transition superoperator%
\begin{equation}
\hat{\Xi}(t)\rho =2bg\left( 1+\bar{n}\right) \sum_{l=0}^{\infty }\left(
2\gamma \bar{n}\right) ^{l}\hat{\Xi}_{l}(t)\rho ,  \label{ttt11}
\end{equation}%
where
\begin{equation}
b\equiv \gamma /g,\qquad \hat{\Xi}_{0}(t)\rho =\hat{\theta}_{0}\rho \hat{%
\theta}_{0}^{\dagger },\quad \hat{\theta}_{0}(t)=\langle e|X_{t}|g\rangle .
\label{ccc}
\end{equation}%
For $l>0$
\begin{equation}
\hat{\Xi}_{l}(t)\rho =\int_{0}^{t}dt_{1}\cdots \int_{0}^{t_{l-1}}dt_{l}\hat{%
\theta}_{l}\rho \hat{\theta}_{l}^{\dagger },\quad \hat{\theta}%
_{l}(t,t_{1},\cdots ,t_{l})=\langle e|X_{t}\widetilde{\sigma }_{+}(t_{1})%
\widetilde{\sigma }_{+}(t_{2})\cdots \widetilde{\sigma }_{+}(t_{l})|g\rangle
.  \label{ttt12}
\end{equation}

As shown in \cite{EQJS} it is enough to evaluate just the three
initial terms in the sum (\ref{ttt11}), whose constituents are found
to be
\begin{gather*}
\hat{\theta}_{0}=-ie^{-\gamma t\left( \overline{n}+1/2\right) -i\omega
\left( \hat{n}+1/2\right) t}S_{\hat{n}+1}(t)\hat{a},\,\qquad \hat{\theta}%
_{1}=e^{-\gamma t\left( \overline{n}+1/2\right) -i\omega \hat{n}t}\chi _{%
\hat{n}+1}(t-t_{1})\chi _{\hat{n}}(-t_{1}), \\
\hat{\theta}_{2}=-ie^{-\gamma t\left( \overline{n}+1/2\right) -i\omega (\hat{%
n}-1/2)t}\chi _{\hat{n}+1}(t-t_{1})S_{\hat{n}}\left( t_{1}-t_{2}\right) \chi
_{\hat{n}-1}(-t_{2})\hat{a}^{\dagger },
\end{gather*}%
where%
\begin{equation}
C_{\hat{n}}(t)=\cos \left( \gamma tB{_{\hat{n}}/b}\right) ,\text{ }S_{\hat{n}%
}(t)=\sin \left( \gamma tB{_{\hat{n}}}/b\right) /{B}_{\hat{n}},\quad \chi _{%
\hat{n}}=e^{-i\omega t/2}\left[ C_{\hat{n}}(t)-i\delta S_{\hat{n}}(t)\right]
,
\end{equation}%
\begin{equation}
{B}_{\hat{n}}=\sqrt{{\hat{n}}+{\delta }^{2}},\quad \delta =(q-ib)/2,\quad
q\equiv \left( \omega _{0}-\omega \right) /g.  \label{vvv13}
\end{equation}

Substituting these expressions into Eq. (\ref{ttt11}), the transition
superoperator turns out to be time-dependent, contrary to the standard
definition of the QJS. So we evaluate the QJS as the time average of $\hat{%
\Xi}(t)$ over the time interval $T$ (to be determined later) during which
the photoelectron is emitted with high probability
\begin{equation}
\hat{J}\rho \equiv \frac{1}{T}\int_{0}^{T}dt\ \hat{\Xi}(t)\rho .
\label{J-Xi}
\end{equation}%
Considering the weak coupling ($\omega ,\omega _{0}\gg \gamma ,g$) for which
the Jaynes--Cummings Hamiltonian (\ref{JCH}) and the master equation (\ref%
{eqmestra}) are valid, and expressing the field density operator in Fock
basis as
\begin{equation}
{\rho }=\sum_{m,n=0}^{\infty }\rho _{mn}|m\rangle \langle n|,
\end{equation}%
after the averaging in (\ref{J-Xi}) the off-diagonal elements of $\hat{J}%
\rho $ vanish due to rapid oscillations of the terms $\exp (\pm i\omega t)$.
Therefore, we are left only with the diagonal elements of $\hat{J}%
\rho $. Applying the superoperators $\hat{\Xi}_{l}$ on the density matrix as
in (\ref{ttt11}) and evaluating Eq. (\ref{J-Xi}) we obtain%
\begin{equation}
\hat{J}\rho =\sum_{n=0}^{\infty }\rho _{nn}\left[
nJ_{n}^{(B)}|n-1\rangle \langle n-1|\,+J_{n}^{(D)}|n\rangle \langle
n|\,+(n+1)J_{n}^{(E)}|n+1\rangle \langle n+1|+\cdots \right] ,
\label{wew}
\end{equation}%
where the explicit expressions for the $n$-dependent functions $J_{n}^{(B)}$%
, $J_{n}^{(D)}$ and $J_{n}^{(E)}$ are given in \cite{EQJS}.

The QJS (\ref{wew}) contains an infinite number of terms, so after a
click the initial field state $\rho $ reduces to a mixture of
different states, each one
with its respective probability. The first term, with coefficient $%
J_{n}^{(B)},$ takes out a photon from the field, so it represents a
click preceded by a photoabsorption -- we call this event a ``bright
count''. The second term, dependent on $J_{n}^{(D)}$ (proportional to $\bar{n}%
,$ quite small as will be shown below), does not subtract photons
from the field but only modifies the relative weight of the field
state components -- it represents a ``dark count'', when the
detector emits a click due to the
amplification of its intrinsic excitations. All further terms in Eq. (\ref%
{wew}) are proportional to $\bar{n}^{l}$, $l\geq 2$; they describe
emissions of several photons into the field upon a click, so we call
the first of these term, $J_{n}^{(E)},$ the ``emission term''. There
are many different phenomena that give rise to dark counts
\cite{Karve,Kitay,Panda}, our model takes into account only those
causing the sensor's ground--excited state transition. Since the
sensor's state depends on its interaction with the field, the dark
counts modify indirectly the relative
weight between the field state components -- that is why they depend on $n$%
, what is not obvious at first glance.

\section{Comparison to experimental data and QJS engineering}

\label{sub2}

Now we compare our results regarding SPD properties with available
experimental data. Experimentally \cite{Verevkin} the dependencies
of bright and dark counts rates are set as functions of the light
wavelength and the detector's ``bias parameter'' (BP). In the BP we
englobe such quantities as bias voltage, bias current and other
physical quantities the experimenter adjusts in order to achieve
simultaneously
the highest Signal-to-Noise ratio $\mathcal{S}$ and bright counts rate. $%
\mathcal{S}\equiv \mathcal{R}_{B}/\mathcal{R}_{D}$ is the ratio between the
bright, $\mathcal{R}_{B}$, and dark, $\mathcal{R}_{D}$, counts rates. When
one increases the BP, the bright counts rate increases while $%
\mathcal{S}$ remains unchanged until the \emph{breakdown} value of BP, when $%
\mathcal{S}$ starts to fall rapidly as function of the BP. So most
detectors usually operate near the BP breakdown in order to achieve
the optimal performance. In practice, $\mathcal{R}_{B}$ is
determined by directing laser pulses containing in average one
photon at a given repetition rate on the detector and measuring the
rate of counts, so in our model it is described by the term
$J_{1}^{(B)}$. Analogously, $\mathcal{R}_{D}$ is calculated as
the rate of counts in the absence of any input signal, so it is given by $%
J_{0}^{(D)}$.
\begin{figure}[tbp]
\includegraphics[width=0.48\textwidth]{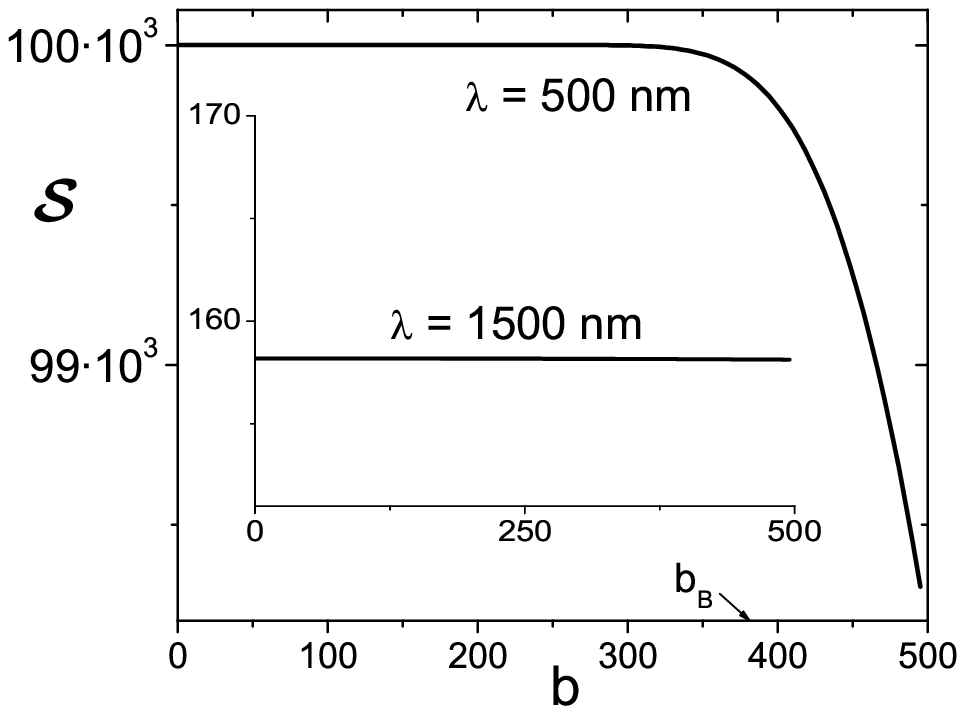} \hfil
\includegraphics[width=0.48\textwidth]{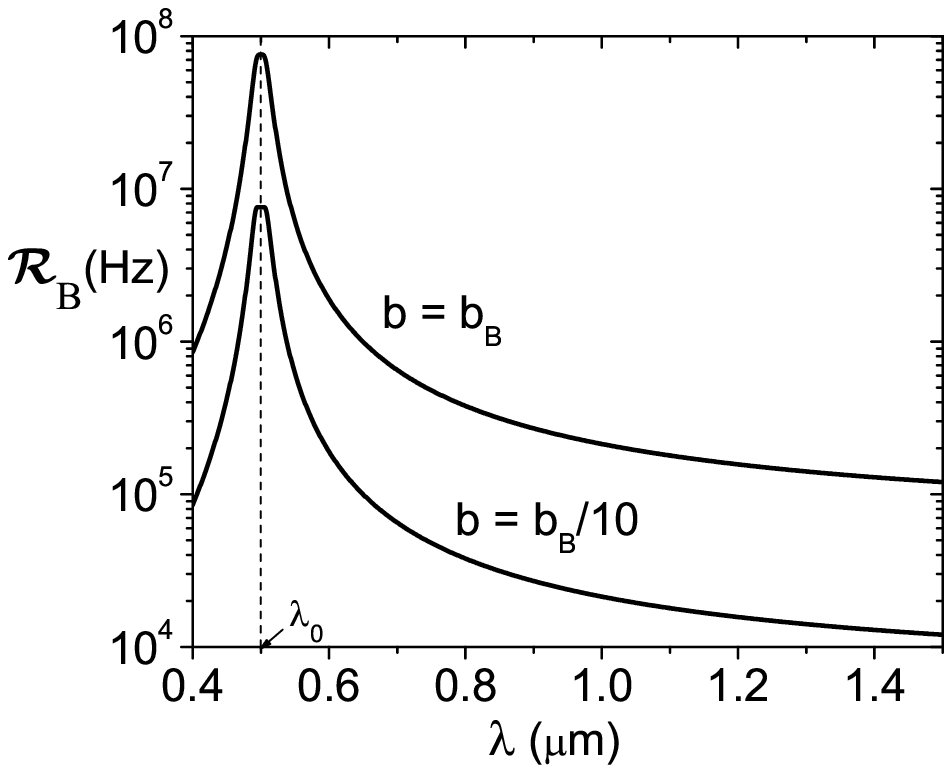} \newline
\parbox[t]{0.48\textwidth}{\caption{Signal-to-Noise ratio as function of $b$ at resonance ($\protect\lambda =500$ nm) and far from it ($\protect\lambda =1500$ nm) in
the inset. The estimated breakdown value is $b_{B}\approx
380$.}\label{fig1}} \hfil
\parbox[t]{0.48\textwidth}{\caption{Bright counts rate as function of wavelength of the field for
different values of $b$, showing that ${\mathcal R}_B$ increases
proportionally to $b$. The resonant wavelength is
$\protect\lambda_0=500$
nm.}\label{fig2}} 
\end{figure}

To do the comparison we need to set the values for our model free
parameters: $\omega _{0}$, $g$, $\Upsilon
=\gamma T$ and $\bar{n}$. For simplicity we shall express the frequencies $%
\omega _{0}$ and $\omega $ in terms of respective wavelengths $\lambda _{0}$
and $\lambda $. Thus we are left with two parameters, $\lambda $ and $b$,
where $b$ plays the role of the BP. Our general model cannot determine the BP%
$\times b$ dependence for every kind of detector; nevertheless, one may
argue that the BP and $b$ must be proportional to each other, since for zero
BP one should also have $b=0$, because in this case the detector would be
turned off. Here, we do not need to know the exact dependence of the BP on $b
$ provided we determine the breakdown value $b_{B},$ corresponding to the BP
breakdown at resonance, and take it as a measure of $b$.

After numerical simulations we have chosen values of the free
parameters that reproduce qualitatively the common experimental
behavior \cite{Kitay,APD1,Verevkin,SSPD} and lie within the
applicability region of the model: $\lambda _{0}=500$ nm,
$g=10^{11}$ Hz, $\Upsilon =5\times 10^{5}$ and $\bar{n}=10^{-11}$,
so $b_{B}\approx 380$, as shown in figure 1. Moreover, we verified
that below $b_{B}$ both $\mathcal{R}_{B}$ and $\mathcal{R}_{D}$ have
approximately linear dependence on $b$, in
agreement with our qualitative arguments. In figure 2 we plot $\mathcal{R}%
_{B}$ for two different values of $b$ as function of the light wavelength,
where we see a good agreement with experimental results \cite{APD1} and can check that $%
\mathcal{R}_{B}$ is proportional to $b$. We also confirmed numerically that $%
\mathcal{R}_{D}$ does not depend on the field wavelength, as
expected.

We verified that for the chosen parameters the emission terms [$%
J_{n}^{(E)}$ and further terms in Eq. (\ref{wew})] are at least 10
orders of magnitude smaller than the dark counts term and even more
for bright counts term in Eq. (\ref{wew}). This confirms that
detectors do not emit photons into the field. Intuitively, the
emission of photons by the detector would be possible only at
temperatures much higher than room temperature through black body
radiation, which is not the case in experiments. Thus, in practice
one is dealing only with bright and dark counts terms that act on
the field simultaneously every time a count is registered, so the
QJS takes the form
\begin{equation}
\hat{J}\rho =\mathrm{diag}\left[ \left( \hat{J}_{B}+\hat{J}_{D}\right) \rho %
\right] ,
\end{equation}%
where diag means diagonal terms in Fock basis. In figure 3 we show the
dependence of the normalized bright counts term $J_{n}^{(B)}/J_{1}^{(B)}$ on
$n$ in di-log scale (for better visualization we joined the points). We note
that near and far away from resonance -- see the values of $\lambda $ in the
caption -- one has nearly polynomial dependence (linear in di-log scale)
\begin{equation}
J_{n}^{(B)}\approx J_{1}^{(B)}n^{-2\beta }=\mathcal{R}_{B}\ n^{-2\beta }
\end{equation}%
with $\beta \approx 1/2$ at resonance and $\beta \approx 0$ far away
from it. Thus, in these cases one can write the operator dependence
of bright counts as
\begin{equation}\label{sasha}
\hat{J}_{B}\rho =\mathcal{R}_{B}\ \left( \hat{n}+1\right) ^{-\beta }\hat{a}%
\rho \hat{a}^{\dagger }\left( \hat{n}+1\right) ^{-\beta },
\end{equation}%
thus recovering the E-model with $\beta \approx 1/2$ at the
resonance and
SD-model with $\beta \approx 0$ far away from it, although the values of $%
\mathcal{R}_{B}$ are different in each case (see figure \ref{fig2}).

\begin{figure}[tbp]
\includegraphics[width=0.48\textwidth]{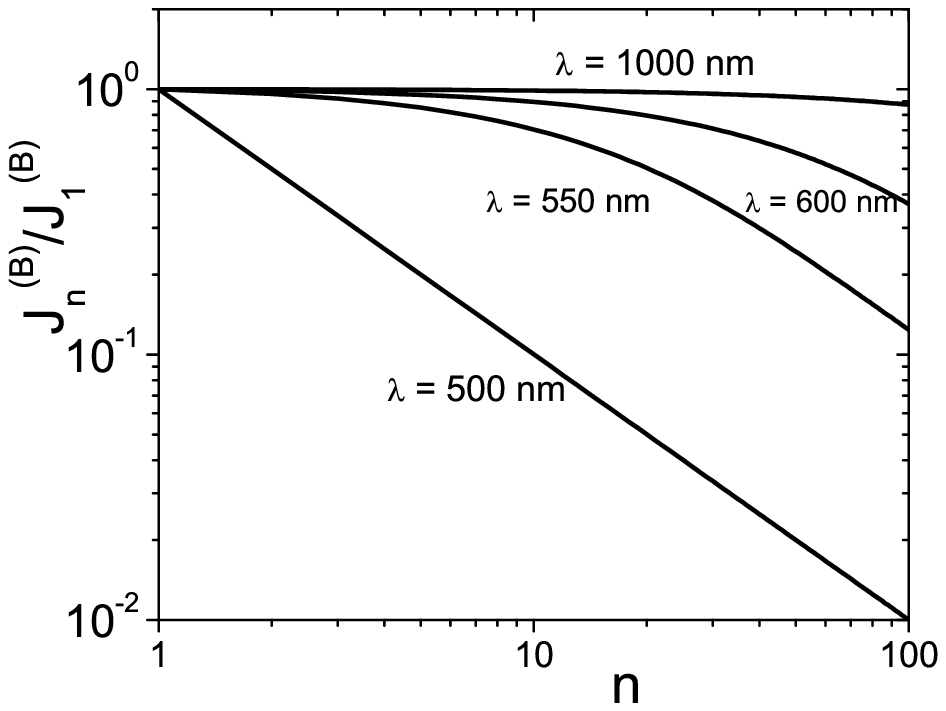} \hfil
\includegraphics[width=0.48\textwidth]{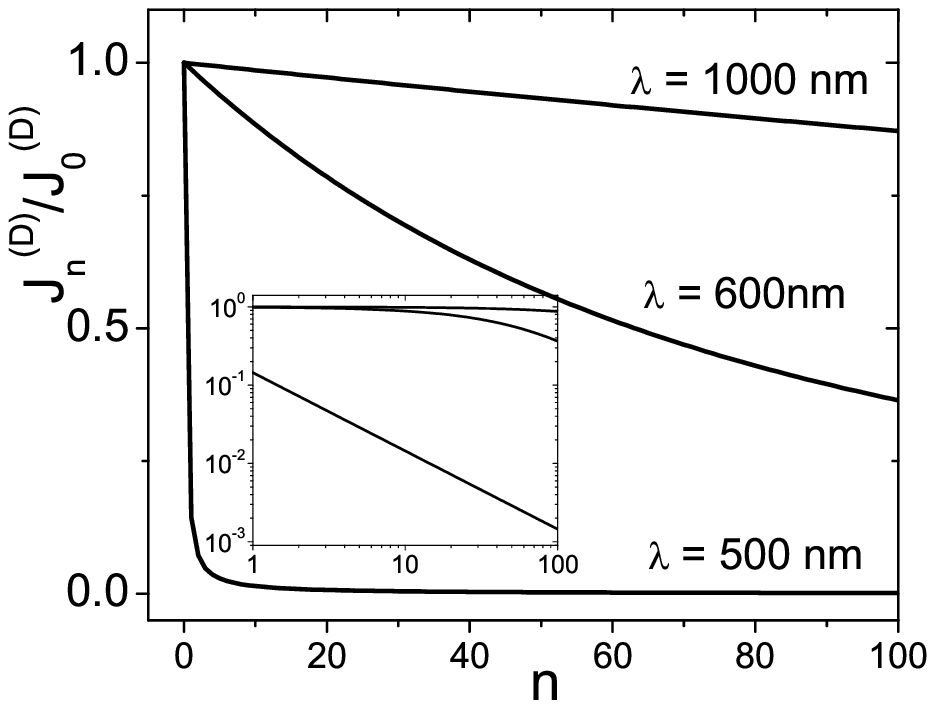} \newline
\parbox[t]{0.48\textwidth}{\caption{Normalized bright counts term as function of $n$ in di-log scale at
breakdown $b_B$. At resonance ($\protect\lambda=500$ nm) we have
$\protect\beta\approx 1/2$ and far away from resonance
($\protect\lambda=1000$ nm) $\protect\beta\approx 0$ [see Eq.
(\ref{sasha})].}\label{fig3}} \hfil
\parbox[t]{0.48\textwidth}{\caption{Normalized dark counts term as function of $n$ at breakdown $b_B$
for different field wavelengths and the same graph in di-log scale
in the inset.}\label{fig4}} 
\end{figure}

The normalized dark counts term $J_{n}^{(D)}/J_{0}^{(D)}$ is shown in figure
4 in linear scale and in di-log scale in the inset. Out of resonance $%
J_{n}^{(D)}$ is almost independent on $n$, so in this case $%
J_{n}^{(D)}\approx \mathcal{R}_{D}=\mathrm{const}$. At the resonance, for $%
n=0$ one gets $J_{0}^{(D)}=\mathcal{R}_{D}$ and for $n>0$ we have $%
J_{n>0}^{(D)}\approx \xi \cdot \mathcal{R}_{D}\cdot n^{-2\beta }$, where $%
\beta \approx 1/2$ and $\xi $ is a number less than $1$
($\mathcal{R}_{D}$ does not depend on the wavelength). This means
that at resonance the dark counts are suppressed in the presence of
light. This happens because they occur when the detector in the
ground state is excited by its intrinsic processes; however, at
resonance the rate of excitations by photons from the field is much
higher than by intrinsic processes, so the dark counts ``have no
time'' to appear. Therefore, the operator form of the dark counts
term in Eq. (\ref{wew}) is
\begin{equation}
\hat{J}_{D}\rho =\mathcal{R}_{D}\left[ \Lambda _{0}\rho \Lambda _{0}+\xi
\Lambda \hat{n}^{-\beta }\rho \hat{n}^{-\beta }\Lambda \right] ,  \label{123}
\end{equation}%
where $\Lambda _{0}\equiv |0\rangle \langle 0|$, $\Lambda \equiv
1-\Lambda _{0}$ and at resonance we have the E-model with $\beta
\approx 1/2$ and $\xi <1$. Far away from resonance we recover the
SD-model with $\beta \approx 0$ and $\xi =1$.

Thus we have exposed our microscopic model for the photodetector and
showed that when one is concerned about the photodetector behavior,
the model agrees with experimental data. Still, the only way to
verify whether the formal expressions of QJSs resulting from the
model hold in practice is to perform photocounting experiments and
compare the outcomes, such as photocounts or WT distributions, to
the model's prediction. In the next sections we shall treat this
issue for realistic situations of detectors with non-unit quantum
efficiency and non-zero dark counts rate. We also discuss possible
measurements that may permit to discern between the E-model and the
SD-model, even in the presence of low efficiency and dark counts.

\section{Including the non-idealities}

\label{sub}

We consider a model for a photodetector with non-unit quantum
efficiency (QE) and a finite dark counts rate. In \cite{VCPM} we
have also considered the effects associated to cavity damping and
the detector's dead-time, and showed that they are not crucial when
compared to the QE and dark counts. Moreover, as the QJS (\ref{01})
is an unbounded superoperator, the inclusion of dead-time effect
into the CPM leads to some inconsistences within SD-model, such as
non-normalizable photodetection distribution. On the other hand, the
E-model is free from such problems.

\subsection{SD-model}

\label{suba}

We consider a free electromagnetic monomodal field of frequency
$\omega $,
enclosed in an ideal cavity together with a photodetector. The \textit{%
unconditioned time evolution} (UTE) of the field in the presence of the
detector, i.e. the evolution when the detector is turned on but the outcomes
of the measurements are disregarded (not registered), is described by the
master equation \cite{milburn1,OMD-JOB,a4}
\begin{equation}
\dot{\rho}=-i\omega \left( \hat{n}\rho -\rho \hat{n}\right) -\frac{\mathcal{R%
}}{2}\left( \hat{n}\rho +\rho \hat{n}-2\hat{A}\rho \right) , \quad \hat{A}\rho \equiv \hat{a}\rho \hat{a}^{\dagger }, \quad \hat{n}=\hat{a%
}^{\dagger }\hat{a}. \label{nc}
\end{equation}%
The first term on the RHS stands for the free field evolution and
the second describes the effect of the detector on the field due to
their mutual interaction. The parameter $\mathcal{R\equiv R}_{B}$
(we omit the subscript $B$ to simplify the notation) is the
field-detector coupling constant, roughly equal to the ideal
counting rate.

To describe photodetection with QE $\eta $ and the dark counts rate
$\mathcal{R}d$ ($d$ is the ratio between the dark counts rate and
the
ideal photon counting rate), we take $\beta =0$ and $\xi =1$ in Eq. (\ref{123}%
)
\begin{equation}
\hat{J}\rho =\mathcal{R}\left( \eta \hat{A}+d\right) \rho .  \label{qjs}
\end{equation}%
The first term within the parenthesis describes the absorption of a
photon from the field with probability per unit time
$\mathrm{Tr}\left[ \eta \mathcal{R}\hat{A}\rho \right] =\eta
\mathcal{R}\bar{n}$, where $\bar{n}$ is the field mean photon number
-- this means that the detector ``sees'' all the photons. The second
term describes the dark counts, so after a detector's click the
field state becomes a mixture of two possible outcomes: either a
photon absorption or a dark count.

The no-count state $\rho _{S}\equiv \hat{S}_{t}\rho _{0}$, where $\hat{S}_{t}
$ is the no-count superoperator, obeys Eq. (\ref{nc}) when one subtracts the
term (\ref{qjs}) on the RHS (see \cite{carmichael,QT11}). Moreover, as we
are interested in calculating probabilities, we shall disregard phase
factors $\exp (\pm i\omega \hat{n}t)$, since they are canceled in the trace
calculation. So the evolution equation of $\rho _{S}$ is
\begin{equation}
\dot{\rho}_{S}=-\frac{\mathcal{R}}{2}\left( \hat{n}\rho _{S}+\rho _{S}\hat{n}%
\right) +\mathcal{R}v\hat{A}\rho _{S}-\mathcal{R}d\rho _{S},\quad v\equiv
1-\eta .  \label{sol}
\end{equation}%
We solved this equation in \cite{VCPM}, obtaining
\begin{equation}
\hat{S}_{t}\rho _{0}=e^{-d\mathcal{R}t}\hat{E}_{t}(e^{v\phi _{t}\hat{A}}\rho
_{0}),\quad \hat{E}_{t}\rho =e^{-\mathcal{R} \hat{n}t/2}\rho e^{-\mathcal{R} \hat{n}%
t/2},\qquad \phi _{t}=1-e^{-\mathcal{R}t}.  \label{st}
\end{equation}%
The field UTE superoperator $\hat{T}_{t}$, defined as the solution to Eq. (%
\ref{nc}), is naturally given by setting $d=\eta =0$ in Eqs.
(\ref{sol}) and (\ref{st}). We introduced in Eq. (\ref{st}) a
compact notation
for the infinite sum in terms of the exponential superoperator:%
\begin{equation}
\exp {(v\phi _{t}\hat{A})}\rho _{0}\equiv \sum_{l=0}^{\infty
}\frac{\left( v\phi _{t}\right) ^{l}}{l!}\hat{a}^{l}\rho _{0}\left(
\hat{a}^{\dagger }\right) ^{l}.
\end{equation}

The $m$-counts superoperator $\hat{N}_{t}(m)$, that describes the field
state after $m$ registered counts (whatever real or dark ones) in the time
interval $(0,t)$, and whose trace gives the probability for this event is%
\begin{equation}
\hat{N}_{t}(m)\rho =\int_{0}^{t}dt_{m}\int_{0}^{t_{m}}dt_{m-1}\cdots
\int_{0}^{t_{2}}dt_{1}\hat{S}_{t-t_{m}}\hat{J}\hat{S}_{t_{m}-t_{m-1}}\hat{J}%
\cdots \hat{J}\hat{S}_{t_{1}}\rho ,  \label{n-nodead}
\end{equation}%
and after some manipulations \cite{VCPM} it reduces to%
\begin{equation}
\hat{N}_{t}(m)=\hat{S}_{t}\frac{(d\mathcal{R}t+\eta \phi _{t}\hat{A})^{m}}{m!%
}.  \label{n12}
\end{equation}%
A simple manner for contrasting the predictions of the model to the
experimental data is by looking to the lower photocounts moments%
\begin{equation}
\bar{m}_{t}=\sum_{m=0}^{\infty }m\mathrm{Tr}\left[ \hat{N}_{t}(m)\rho _{0}%
\right] =d\mathcal{R}t+\eta \bar{n}\phi _{t}  \label{msd}
\end{equation}%
\begin{equation}
\overline{m(m-1)}_{t}=\left( d\mathcal{R}t\right) ^{2}+2\eta \bar{n}d%
\mathcal{R}t\phi _{t}+(\eta \phi _{t})^{2}\overline{n(n-1)},
\end{equation}%
where $\bar{n}$ and $\overline{n(n-1)}$ are the factorial moments of the
initial density operator.

Another measurable quantity we consider is the \textit{waiting time
distribution}. It describes the probability density for registering two
consecutive clicks separated by the time interval $\tau $, provided the
first one occurred at time $t$. Its non-normalized form is%
\begin{equation}
\mathrm{W}_{t}(\tau )=\mathrm{Tr}\left[ \hat{J}\hat{S}_{\tau }\hat{J}\hat{T}%
_{t}\rho \right] ,  \label{www}
\end{equation}%
and the mean WT is%
\begin{equation}
\bar{\tau}=\mathcal{N}^{-1}\int_{0}^{\theta }d\tau \mathrm{W}_{t}\left( \tau
\right) \tau ,\quad \mathcal{N}=\int_{0}^{\theta }d\tau \mathrm{W}_{t}\left(
\tau \right) ,
\end{equation}%
where $\theta $ is the time interval during which one evaluates the
averaging in experiments. As to be shown in section \ref{sub3},
$\theta $ is an important parameter due to the presence of dark
counts. We obtained the expression for ${\mathrm{W}}_{t}(\tau )$ in
\cite{VCPM}:
\begin{equation}
{\mathrm{W}}_{t}(\tau )=e^{-d\mathcal{R}\tau }\left\{ \eta ^{2}e^{-\mathcal{R%
}\left( 2t+\tau \right) }\Phi _{2}^{W}+\eta de^{-\mathcal{R}t}(1+e^{-%
\mathcal{R}\tau })\Phi _{1}^{W}+d^{2}\Phi _{0}^{W}\right\} ,
\end{equation}%
where
\begin{equation}
\Phi _{k}^{W}=\sum_{n=k}^{\infty }\rho _{n}\frac{n!}{(n-k)!}\left( 1-\eta
\phi _{\tau }e^{-\mathcal{R}t}\right) ^{n-k},\qquad \rho _{n}=\langle n|\rho
|n\rangle .  \label{lordi}
\end{equation}

\subsection{E-model}

\label{subb}

We now repeat the same procedures for the E-model with the QJS
\begin{equation}
\hat{J}\rho =\mathcal{R}\left( \eta \hat{\varepsilon}+d\right) \rho ,\qquad
\hat{\varepsilon}\rho \equiv \hat{E}_{-}\rho \hat{E}_{+}.
\end{equation}%
For simplicity we considered a simplified form for the dark counts term,
analogous to the one we used in SD-model, but different from the one given
by Eq. (\ref{123}). The probability per unit time for detecting a photon is $%
\eta \mathcal{R}(1-p_{0})$, where $p_{0}=\langle 0|\rho |0\rangle $,
so the detector ``sees'' whether there is a photon in the cavity. In
principle, the parameter $\mathcal{R}$ is different from the one in
SD-model, but here it will be always clear which one we are dealing
with. The field UTE is described by an equation similar to Eq.
(\ref{nc}), obtained by doing the
substitution $\left\{ \hat{a},\hat{a}^{\dagger }\right\} \rightarrow \{\hat{E%
}_{-},\hat{E}_{+}\}$ in the non-unitary evolution [second term on
the RHS of Eq. (\ref{nc})]. So the no-count state $\rho _{S}$ obeys
the equation
\begin{equation}
\dot{\rho}_{S}=-\frac{\mathcal{R}}{2}\left( \hat{\Lambda}\rho _{S}+\rho _{S}%
\hat{\Lambda}\right) +\mathcal{R}v\hat{\varepsilon}\rho _{S}-d\mathcal{R}%
\rho _{S},  \label{hgf}
\end{equation}%
[similar to Eq. (\ref{sol})] where $\hat{\Lambda}\equiv \hat{E}_{+}\hat{E}%
_{-}=1-\hat{\Lambda}_{0}$, $\hat{\Lambda}_{0}\equiv |0\rangle \langle 0|$.
Since we are going to calculate probabilities, it is sufficient to write out
just the \emph{diagonal} form of the no-count superoperator in the Fock
basis, given by \cite{VCPM}%
\begin{equation}
\hat{S}_{t}\cdot =e^{-d\mathcal{R}t}\ \left[ \hat{P}_{t}\cdot +\hat{\Lambda}%
_{0}\frac{1-\hat{P}_{t}}{1-v\hat{\varepsilon}}\cdot \hat{\Lambda}_{0}\right]
,\qquad \hat{P}_{t}\equiv e^{-\mathcal{R} t\left( 1-v\hat{\varepsilon}%
\right) },  \label{ste1}
\end{equation}%
where the dot $\cdot $ stands for any density operator. Once again, the
functions of the superoperator $\hat{\varepsilon}$ should be calculated as
power series.

The $m$-counts superoperator is%
\begin{eqnarray}
\hat{N}_{t}(m)\cdot  &=&e^{-d\mathcal{R}t}\left\{ \left( 1-\hat{\Lambda}_{0}%
\frac{1}{1-v\hat{\varepsilon}}\cdot \hat{\Lambda}_{0}\right) \hat{P}_{t}%
\frac{(\hat{J}t)^{m}}{m!}+\hat{\Lambda}_{0}\frac{1}{1-v\hat{\varepsilon}}%
\frac{\left( d\mathcal{R}t\right) ^{m}}{m!}\cdot \hat{\Lambda}_{0}\right.
\notag \\
&&\left. +\hat{\Lambda}_{0}\frac{\mathcal{R}\eta \hat{\varepsilon}}{1-v\hat{%
\varepsilon}}\int_{0}^{t}dx\hat{P}_{x}\frac{\left[ d\mathcal{R}t+\eta \hat{%
\varepsilon}\mathcal{R}x\right] ^{m-1}}{(m-1)!}\cdot \hat{\Lambda}%
_{0}\right\} ,
\end{eqnarray}%
where the last term is zero for $m=0$, and the expressions for the initial
factorial photocounts moments read%
\begin{equation}
\overline{m}_{t}=d\mathcal{R}t+\eta \bar{n}\left( 1-\Xi _{1}\right) ,
\label{me}
\end{equation}%
\begin{equation}
\overline{m(m-1)}_{t}=\left( d\mathcal{R}t\right) ^{2}+2\eta \bar{n}d%
\mathcal{R}t\left( 1-\Xi _{1}\right) +\eta ^{2}\left[ \overline{n(n-1)}%
\left( 1-\Omega \right) -2\bar{n}\mathcal{R}t\Xi _{2}\right] ,  \label{me2}
\end{equation}%
where%
\begin{equation}
\Xi _{k}\equiv \frac{1}{\bar{n}}\mathrm{Tr}\left[ \frac{\hat{\varepsilon}^{k}%
}{1-\hat{\varepsilon}}\hat{P}_{t}^{0}\rho \right] ,\qquad \Omega \equiv
\frac{2}{\overline{n(n-1)}}\mathrm{Tr}\left[ \left( \frac{\hat{\varepsilon}}{%
1-\hat{\varepsilon}}\right) ^{2}\hat{P}_{t}^{0}\rho \right] ,\quad \hat{P}%
_{t}^{0}\equiv \hat{P}_{t}(v=1).  \label{Chik}
\end{equation}%
Using Eq. (\ref{www}), the WT distribution is found to be
\begin{equation}
\mathrm{W}_{t}\left( \tau \right) =e^{-d\mathcal{R}\tau }\left\{ (\mathcal{R}%
d)^{2}[1-\mathrm{Tr}(\hat{P}_{t}^{0}\rho )]+\mathrm{Tr}[(\hat{J}\hat{P}%
_{\tau }+\mathcal{R}d\hat{\Lambda}_{0}\frac{1-\hat{P}_{\tau }}{1-v\hat{%
\varepsilon}}\cdot\hat{\Lambda}_{0})\hat{J}\hat{P}_{t}^{0}\rho
]\right\} . \label{WE}
\end{equation}

\section{Some schemes for verifying CPM}

\label{sub3}

Guided by experimental data \cite{had} we chose the following numerical
values for the model parameters: $\eta =0.6$ for the QE and $d=5\cdot 10^{-3}
$ for the dark counts rate (normalized by the ideal counting rate). We do
not attribute any fixed value to $\mathcal{R}$ since our analysis will be
given in terms of the dimensionless $\mathcal{R}t$ ($t$ being the time). As
many photodetection quantities were reported in different contexts  \cite%
{SD,ueda2,a17,SD1,OMD-JOB,waiting1,waiting4,DMD-JOB05}, we shall
consider few of them that could help to decide between the SD- or
the E- model.

\begin{figure}[tbp]
\includegraphics[width=0.48\textwidth]{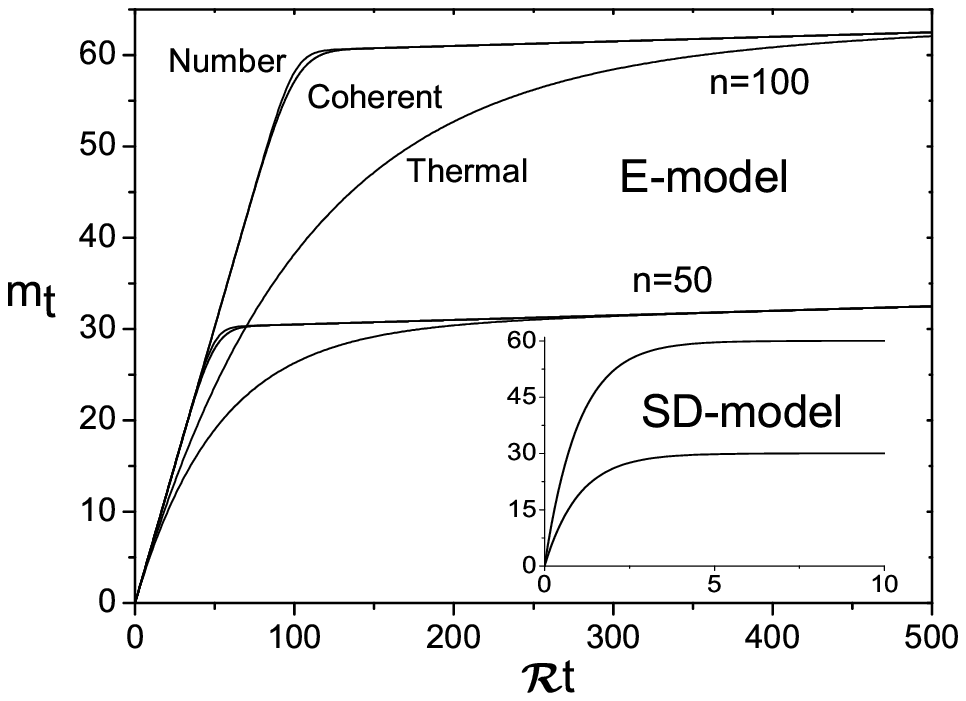} \hfil
\includegraphics[width=0.48\textwidth]{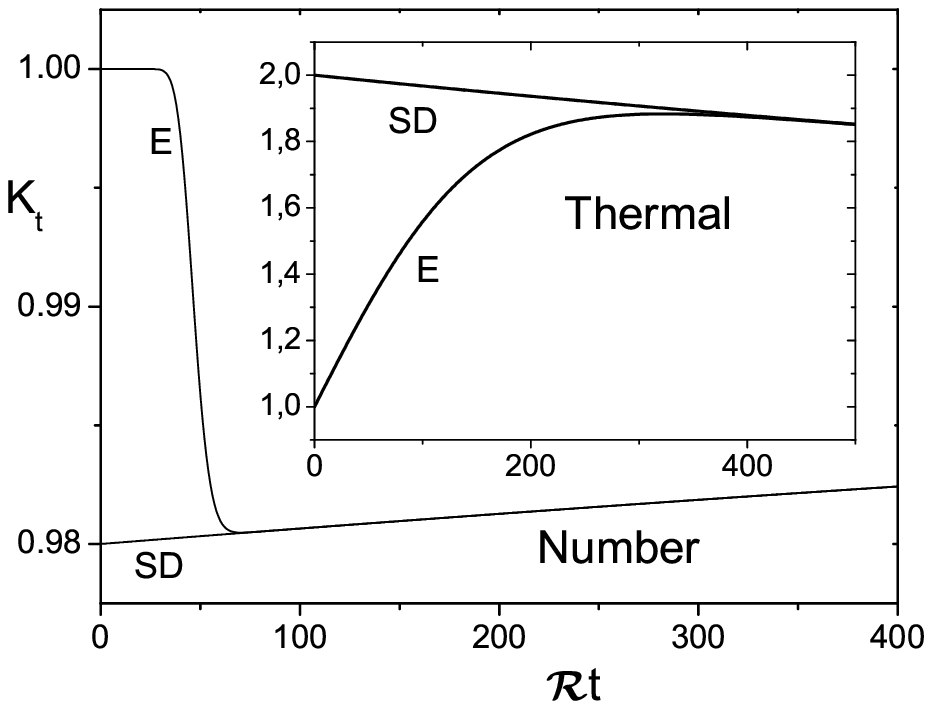} \newline
\parbox[t]{0.47\textwidth}{\caption{$\bar{m}_{t}$ in E-model for coherent,
number and thermal states (indicated in the figure, the lower curves
are labeled analogously) as function of time: the lower curves
correspond to $\bar{n}=50$ and the upper -- to $\bar{n}=100$. In the
inset we plot $\bar{m}_{t}$ for the SD-model, which is independent
from field state.}\label{figure1}} \hfil
\parbox[t]{0.47\textwidth}{\caption{$K_{t}$, Eq. (\protect\ref{kkk}), for SD- and E- models (as indicated in the graph with
abbreviations) for the number state (and the thermal state in the
inset) for $\bar{n}=50$. For the coherent state one has $K_{t}=1$ at
all times for both models.}\label{figure2}} 
\end{figure}

We first analyze the counting statistics. In figure \ref{figure1} we plot $%
\bar{m}_{t}$ as function of $\mathcal{R}t$ for both models for two values of
the initial mean photon number, $\bar{n}=50$ and $100$. Initially, $\bar{m}%
_{t}$ increases steeply due to photons absorption, and after some
time the growth turns linear with much smaller slope due to the dark
counts. We call the time interval during which the photons are
absorbed (representing the
duration of the steep increase in the number of counts) the \textit{%
effective counting time}\emph{\ }$t_{E}$. In the E-model $t_{E}$ is
proportional to the initial average photon number, contrary to the SD-model
[as seen from the figure \ref{figure1} and formulae (\ref{msd}) and (\ref{me}%
)]. So the experimental analysis of the dependence of $t_{E}$ on $\bar{n}$
seems to us a feasible way for verifying which model could hold in practice,
because, according to the SD-model, $t_{E}$ does not depend on $\bar{n}$.
Moreover, one could also check the validity of each model by verifying
whether $\bar{m}_{t}$ depends on the initial field state: in the SD-model it
is independent of the field state, while in the E-model $\bar{m}_{t}$ is
quite sensible to it: in figure \ref{figure1} one sees a notable difference
between thermal and coherent states, although not so much between number and
coherent states. This can be explained by a great difference in the values
of Mandel's $Q$-factor \cite{qfactor} characterizing the statistics of
photons in the initial state: it equals $-1$ and $0$ for number and coherent
states, respectively, whereas it is very big ($Q_{th}=\bar{n}$) for the
thermal states with big mean numbers of photons.

Now we analyze the normalized second factorial moment
\begin{equation}
K_{t}\equiv {\overline{m(m-1)}_{t}}/{\overline{m}_{t}^{2}}
\label{kkk}
\end{equation}%
for the same initial states with mean photon number $\bar{n}=50$. For the
number and thermal states $K_{t}$ as function of $\mathcal{R}t$ is shown in
figure \ref{figure2}, and for the coherent state we get $K_{t}=1$, so it is
not plotted. In the asymptotic time limit and for non-zero dark counts rate,
the same value $K_{\infty }\rightarrow 1$ holds for both models, however the
transient is model dependent. In the SD-model without considering dark
counts $K_{t}$ is time-independent, $K_t=\overline{n(n-1)}/\overline{n}^{2}$ ($%
\bar{n}$ and $\overline{n(n-1)}$ correspond to the initial field
state), nevertheless it depends on the initial field state: $K_t=2$
for the thermal state and $K_t=1-1/\bar{n}$ for the number state. By
including the dark counts in the analysis this constant behavior is
slowly modified as time goes on, see figure \ref{figure2}. In the
E-model in the absence of dark counts $K_{t} $ starts at the value
\begin{equation}
\lim_{t\rightarrow 0}K_{t}=\frac{\mathrm{Tr}\left( \hat{\varepsilon}^{2}\rho
\right) }{\left[ \mathrm{Tr}\left( \hat{\varepsilon}\rho \right) \right] ^{2}%
}=\frac{1-\rho _{0}-\rho _{1}}{\left( 1-\rho _{0}\right) ^{2}},
\end{equation}%
which is exactly $1$ for the number state and very close to $1$ for the
thermal state with the chosen values of $\bar{n}$. With the course of time, $%
K_{t}$ attains the same values as for the SD-model (for respective
initial field states) when all the photons have been counted. By
taking in account the dark counts, such a behavior is slightly
modified, yet it is quite
different from the behavior in the SD-model, as shown in the figure \ref%
{figure2}. This is another possible manner for verifying the applicability
of SD- or E- models.

We now turn our attention to the WT analysis. It is important to
define the time interval over which we do the average: if one has
non-zero dark counts rate, then by performing the average over a
very large time
interval, we shall always get for the mean WT the value $\bar{\tau}%
\sim \left( \mathcal{R}d\right) ^{-1}$, which is nothing but the
mean time interval between consecutive dark counts. Since
experimentally the average is done over finite time intervals, we
shall proceed in the same way: the mean WT for initial times, when
the photon number is
significative, is roughly $(\eta \mathcal{R})^{-1}$ (because $\eta \mathcal{R%
}$ is the effective counting rate), so we shall take the average
over a time interval $\theta =10\,(\eta \mathcal{R})^{-1}$. This
means that if one does not detect consecutive counts within the time
$\theta $, such a measurement will not contribute to the average. In
an ideal case this procedure is not necessary because the
probability for registering consecutive clicks separated by a large
time interval is zero.

\begin{figure}[h]
\includegraphics[width=.48\textwidth]{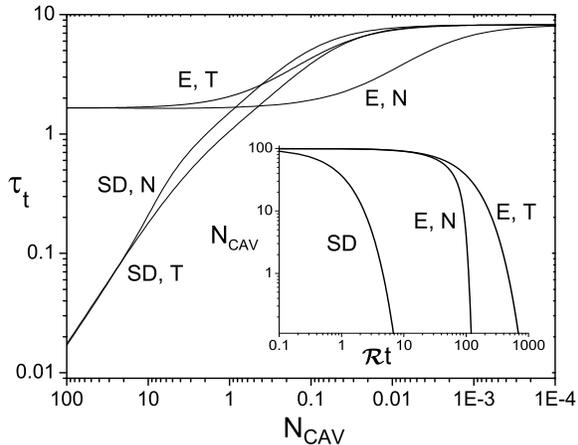}\hspace{2pc}%
\begin{minipage}[b]{18pc}\caption{\label{figure3}Mean WT $\bar{\protect\tau}_{t}$ as function of $N_{CAV}$
for the number (N) and thermal (T) states for SD- and E- models.
While there are photons in the cavity $\bar{\protect\tau}_{t}$ is
constant for the
E-model, but increases substantially with time for the SD-model. In the inset we plot $%
N_{CAV}$ as function of $\protect{\mathcal R} t$ (assuming the same
${\mathcal R}$ for both models) for these states (in the SD-model
$N_{CAV}$ is state independent).}
\end{minipage}
\end{figure}

In figure \ref{figure3} we plot the mean WT for the SD- and E-
models, for the number and thermal initial states (for the coherent
state we
obtain a curve almost identical to the one for the number state) with $\bar{n}%
=100$ as function of the mean photon number in the cavity at the moment of
the first click,
\begin{equation}
N_{CAV}=\mathrm{Tr}\left[ \hat{n}\hat{T}_{t}\rho _{0}\right]=\left\{
\begin{array}{c}
\bar{n}e^{-{\mathcal R}t}\mbox{ for SD-model} \\
\bar{n}\Xi
_{1}\mbox{ for E-model.}%
\end{array}%
\right.  \label{ncav}
\end{equation}%
(For completeness, in the inset of figure \ref{figure3} we plot
$N_{CAV}$ as function of $\mathcal{R}t$ for both models.) For the
E-model, we see that when $N_{CAV}$ becomes less than $1$, the WT
starts to increase visibly due to the dominance of dark counts,
which are much more rare events than absorption of photons. This is
a drastic departure from the ideal case, in which no counts occur
after all the photons have been absorbed, so the mean WT saturates
at the inverse value of the
counting rate, as shown in \cite{DMD-JOB05}. Moreover, from figure \ref%
{figure3} one verifies that as long as there are photons in the
cavity the mean WT is nearly time-independent within the E-model
(and truly independent in the ideal case \cite{DMD-JOB05}), and does
increase substantially in time for SD-model. This is another notable
qualitative difference that could be verified experimentally.

\section{Summary and conclusions}

\label{secc}

We presented a microscopic model for a photodetector modeled as a
2-level quantum sensor plus a macroscopic amplification mechanism.
Using the quantum trajectories approach we deduced a general QJS
describing the back-action of the detector on the field upon a
photocount and showed that it can be represented formally as an
infinite sum of terms. In that sum we have identified the terms
corresponding to the bright counts (photoabsorptions), the dark
counts and emission events, each one occurring with its respective
probability. Adjusting the free parameters of the model to fit
experimental data, we showed that the emission terms can be
disregarded in realistic situations since their contribution becomes
insignificant, so the QJS consists effectively only of bright and
dark counts terms. Moreover, we have simulated the experimental
behavior of the counting rates and the Signal-to-Noise ratio,
showing the breakdown phenomenon. We have also showed that with the
detector operating near its breakdown bias one can engineer the QJS
by modifying the wavelength of the field. In particular, one
recovers the QJSs proposed previously \textit{ad hoc}: at resonance
one gets the E-model, and far away from it the SD-model is
identified.

We have also generalized the continuous photodetection model through
a quantum treatment of non-ideal effects that are ubiquitous in
experiments. We derived general expressions for the fundamental
operations in the presence of non-unit quantum efficiency and dark
counts, and calculated explicitly the photocounts and the waiting
time probability distributions for initial coherent, number and
thermal field states. By calculating the first and second factorial
moments of the photocounts and the mean waiting time, we showed that
in standard photodetection experiments one could check the
applicability of the QJS of SD- or E- models. Namely, we indicated
three different ways for revealing the actual QJS: (1)
quantitatively, by studying the time dependence of the normalized
second factorial photocounts moment. Qualitatively, we showed that
the models can also be distinguished by measuring: (2) whether the
effective detection time depends on the initial average photon
number in the cavity and (3) whether the mean waiting time is
modified as time goes on. Still, if the experimental data would
depart significantly from the theoretical predictions one should
reconsider both models and try to look for alternative mechanisms to
reproduce the outcomes.

\begin{acknowledgments}
Work supported by FAPESP (SP, Brazil) contract \# 04/13705-3. SSM
and VVD acknowledge partial financial support from CNPq (DF,
Brazil). \end{acknowledgments}

\end{document}